\documentclass[12pt]{iopart}

\usepackage{graphicx}
\usepackage{dcolumn}

\newcommand{\vect}[1]{{\bf #1}}

\newcommand{\beq}{\begin{equation}}
\newcommand{\eeq}{\end{equation}}
\newcommand{\beqar}{\begin{eqnarray}}
\newcommand{\eeqar}{\end{eqnarray}}

\begin{document}
\title{Collapsing the Wavefunction of a Baseball}
\author{Tarun Biswas}
\address{State University of New York at New Paltz, \\ New Paltz,  NY 12561, USA.}
\ead{biswast@newpaltz.edu}
\begin{abstract}
The transition from microscopic to macroscopic in quantum mechanics can be
seen from various points of view. It is often not merely a transition from quantum
to classical mechanics in the sense of the Correspondence Principle. The fact that
real macroscopic objects like baseballs are composites of an extremely large
number of microscopic particles (electrons, protons etc.) is a complicating
factor. Here such composite objects are studied in some detail and the Copenhagen 
interpretation is applied to them. A computer game model for a simplified composite 
object is used to illustrate some of the issues.
\end{abstract}
\pacs{01.50.-i, 01.50.Wg, 03.65.Ta}
\section{Introduction}
Since the advent of quantum physics about a century ago, physicists have
valiantly tried to understand it at an intuitive level. However, our normal intuition
is developed and exercised almost exclusively in the classical world of
macroscopic objects. So any attempt at understanding quantum mechanics
intuitively is predisposed to fail. Nonetheless, we keep trying in the hope
of at least partial success.

I remember one such attempt in my first quantum physics course in college.
We learnt that the longer the wavelength of a wave, the easier it is to detect its wave
nature (using diffraction experiments). Diffracting sound waves is almost
trivial, but for light waves we need one or more extremely narrow slits
(gratings with spacing of the order of 10$^{-6}$ m). For X-rays and electrons 
the grating pattern has to be finer still (of the order 10$^{-10}$ m
as found in crystals). Now, what about the diffraction of a ``beam'' of 
baseballs? Using the de Broglie formula, for a nominal baseball momentum,
one computes slit widths needed to diffract such a beam to be of the order
of 10$^{-34}$ m! The impossibility of such a grating is given as the reason 
why baseballs are not seen to diffract.

However, a closer look at such reasoning shows serious flaws. If slit widths
of the order of 10$^{-34}$ m were actually possible, there is still no way one
can imagine baseballs getting through them (baseballs are simply too big!). 
The transition
from an electron to a baseball is not merely a change in mass. An electron
is a structureless fundamental particle whereas a baseball 
is a composite of an enormous number of fundamental particles bound 
together. This means that an electron has no measure of spatial extent
or size but the baseball has a well defined measure of size based on equilibrium
distances between component particles. For the baseball to squeeze through
a narrow slit, the distances between its component particles must be decreased 
dramatically. This would require ridiculously high energies. The electron being
structureless will have no problems squeezing through any slit. Hence, the composite
nature makes the baseball qualitatively very different. A proton is a composite
of only three objects and already it shows a magnetic moment and scattering
cross sections (Bjorken scaling) that are very different from structureless particles.

So, here I shall take a closer look at the quantum mechanics of composite objects.
In particular, the process of position observation and the resulting collapse of
wavefunctions of composites like baseballs will be studied. For this
purpose, a sharpened form of the Copenhagen interpretation will be used. 
Also, a computer game based on these ideas will be used to provide some of that
illusive ``intuitive'' understanding.

\section{Observing a baseball}
When the catcher observes the position of a baseball, it is {\em not} a direct
measurement of the center of mass (CM). Hence, the baseball wavefunction
does {\em not} collapse to an eigenstate of CM position. Consequently,
the momentum does {\em not} become particularly uncertain\footnote{Apologies 
are in order for ruining many physics party jokes 
based on this idea of baseball wavefunction collapse.}. To verify these statements, 
let us look at 
this process of baseball position observation in some detail.

When the catcher ``sees'' the baseball, he/she makes myriads of observations --
mostly energies and momenta of photons scattered from the surface of the
ball\footnote{There may also be some photons emitted by excited surface electrons.
These photons carry information about electron energy states. I shall not discuss
such photons here as they are not primarily responsible for position observation.}.
Photon energies are determined by the colors observed. The colors
also help distinguish the ball from its background thus allowing the catcher
to gaze in the direction of the ball. This direction gives the directions of
the photon momenta. The magnitudes of photon momenta are known from the colors.
Hence, the photon momenta are known both in magnitude and direction. 
The momentum of a specific
photon locates the position of the surface electron that scattered it. This kind of
position measurement of different components of the surface of the ball
allows the estimation of the CM position without actually measuring it.
Thus total momentum does not become unduly uncertain.

The above explanation is significantly different from that offered in standard 
introductory courses where the uncertainty in CM position is estimated
to be the diameter of the ball. Then the uncertainty in total momentum
is computed using the uncertainty principle. This also shows that total momentum
uncertainty is not alarmingly large.

\section{Keeping up with Copenhagen}
In the last section it was argued that CM position is estimated from position
measurements of surface particles (mostly electrons). However, even surface
electron positions are not being measured directly. They are measured through
the photons they scatter. So, should the wavefunction of such a surface electron
collapse?

There may be some controversy about this question. Some (following Wigner) give the
Copenhagen interpretation a somewhat anthrocentric tilt. They say that wavefunction 
collapse is achieved only by an observation made by a conscious 
observer\cite{Griffiths,Jammer,Wigner} and as no conscious
observer observes those surface electrons directly, their wavefunctions do not
collapse. However, this understanding 
can quickly drag physics into metaphysics.
The contentious definition of consciousness becomes central to the argument.

On the other hand, if we were to consider an observer to be just a large
collection of particles capable of recording data by selecting particle
states (brain chemistry)\cite{Griffiths,Schrod}, our understanding can be based on known physics.
So, the observer is a macroscopic recorder that stores information
in multiple particle states (often redundantly) as a consequence of an
observation. The photon scattered from an electron is a microscopic 
analog of the same thing\cite{Griffiths,Heisenberg}. It changes its own state
due to the scattering and hence, records information about the scattering
agent (the electron). Thus the scattering of the photon is a measurement
of electron position and must result in the collapse of the electron
wavefunction.

Getting back to the baseball, we see that the indirect estimation of the
baseball position requires the collapse of wavefunctions of a negligibly
small number of surface electrons. Hence, for all practical purposes, the
baseball is unaffected by the measurement. This justifies the classical
model of the baseball. The few electrons that suffer wavefunction collapse
develop high momentum uncertainties and may even escape the binding forces
that attach them to the ball. The computer model discussed later shows this
effect.

\section{Dynamics of composite objects}
Macroscopic objects like baseballs have a very large number of degrees of
freedom due to their large number of component particles. When
the position of one of these particles (say an electron) is measured, the 
whole baseball wavefunction collapses to an eigenstate of position of just
that one electron for a specific position eigenvalue. However, such a
baseball eigenstate must be highly degenerate as its dependence on
all other particle positions can be anything. The collapsing process
needs to select one of these many degenerate states. To understand this
we need to study the dynamics of a composite object with a large number of component 
particles.

The simplest composite would be a bound system of two particles. This
system has been studied extensively both classically and quantum mechanically
as seen in introductory texts\cite{Goldstein,Shankar,Biswas}. However, the two-particle
case is a special one. For most realistic interaction potentials, this
can be separated into two independent one-particle problems -- one a
free particle (CM position) and the other a particle interacting with
a background potential.

A similar separation for three or more interacting particles is unknown. This 
makes analytical solutions for such systems difficult\footnote{If the particles
interact only in pairs, such separation is of course possible. Anything
more than pair interactions can be introduced as perturbations if they are
small enough. For example, Lagrange points in planetary motion are found
in this manner\cite{Whittaker}. For quantum systems, Hartree-Fock type of methods
are of a similar nature.}. However, numerical
solutions are quite straightforward (although time consuming). Hence, I shall
outline such a numerical approach for the non-relativistic quantum 
case\footnote{The relativistic many particle case becomes significantly more 
complicated. To
reconcile relativity and quantum mechanics for arbitrary numbers of particles,
second quantized field theories are needed in general. However, in some special cases
only limited second quantization is enough\cite{Biswas2,Biswas3}.}. 
It can be implemented on a PC with
some approximations. This implementation, in spite of the approximations,
provides physical insight into the wavefunction collapse process. It can
simulate component particle detection by mouse clicks (turning the
simulation into a game).

Let the composite object consist of $N$ mutually interacting 
distinguishable\footnote{In general, a macroscopic object will have a large
variety of component particles -- some distinguishable and some indistinguishable.
Here, for simplicity, I choose them all to be distinguishable.} particles with
positions $\vect{q}_{i}$ and momenta $\vect{p}_{i}$ ($i$=1, 2, \ldots, $N$).
Let the hamiltonian for this object be $H(\{\vect{q}_{i}\},\{\vect{p}_{i}\})$
and its wavefunction at time $t$ be $\psi(\{\vect{q}_{i}\},t)$. Then the integral
form of the time dependent Schr\"{o}dinger equation gives the time development
of the wavefunction to be
\beq
\psi(\{\vect{q}_{i}\},t) = e^{-iHt/\hbar}\psi(\{\vect{q}_{i}\},0). \label{eq1}
\eeq
While no observations are made on the object, the above equation provides
the wavefunction at anytime. It can also be written as
\beq
\psi(\{\vect{q}_{i}\},t) = \lim_{\Delta t\rightarrow 0}
(1-iH\Delta t/\hbar)^{n}\psi(\{\vect{q}_{i}\},0), \label{eq1aa}
\eeq
where $n$ is an integer and $n\Delta t = t$. For a numerical evaluation, 
$\Delta t$ can be chosen
to be small but finite. Then the time development of $\psi$ can be computed
in time steps of $\Delta t$. Using equation~(\ref{eq1aa}), it can be seen that
the values of $\psi$ at successive time steps are
related as follows\footnote{It is to be noted that for $\Delta t\rightarrow 0$, 
this becomes the differential form of the Schr\"{o}dinger equation.}.
\beq
\psi(\{\vect{q}_{i}\},t) = (1-iH\Delta t/\hbar)\psi(\{\vect{q}_{i}\},t-\Delta t). \label{eq1a}
\eeq
This can be used for an iterative computation of $\psi$ if the initial value is
known. Although this method becomes unstable for large numbers of time steps\cite{Teukolsky},
it is adequate for the present application. Here, after every few steps, the wavefunction 
will be collapsed. This prevents the solution from becoming unstable.

An iterative computation using short
time steps also allows computer animation of the time development. The
wavefunction, at each time step, provides a frame for animation.

If a particle detector detects the $k^{\mbox{th}}$ particle in a small region
$R$, it will be with a probability $P_{k}$ given by

\beq
P_{k} = \int_{R}d^{3}\vect{q}_{k}\int_{-\infty}^{\infty}\int_{-\infty}^{\infty}
\cdots\int_{-\infty}^{\infty}\prod_{j\neq k}d^{3}\vect{q}_{j}
\psi^{*}(\{\vect{q}_{i}\},t)\psi(\{\vect{q}_{i}\},t). \label{eq2}
\eeq

If the particle is actually detected, the wavefunction must collapse to
\beq
\psi_{c}(\{\vect{q}_{i}\},t)=A\Delta_{R}(\vect{q}_{k}-\vect{q}_{0})
                         \psi(\{\vect{q}_{i}\},t), \label{eq3}
\eeq
where $\Delta_{R}(\vect{q}_{k}-\vect{q}_{0})$ is a sharply peaked
function that is nonzero only in the region of detection $R$ centered
about the position $\vect{q}_{0}$. The detailed form of this function
depends on the detector sensitivity in the region $R$. In the limit
$R\rightarrow 0$, it is the Dirac delta function:
\beq
\lim_{R\rightarrow 0}\Delta_{R}(\vect{q}_{k}-\vect{q}_{0})
      =\delta^{3}(\vect{q}_{k}-\vect{q}_{0}). \label{eq4}
\eeq
The constant $A$ is needed to renormalize $\psi$ after the collapse.
It is to be noted that the collapsed wavefunction is highly
degenerate because only one particle position is measured and the other particles
could be in many different states without changing this measurement. So one
needs to select a linear combination of these degenerate states to be $\psi_{c}$. This is
done by setting the amplitudes of the members of the degenerate set in
the same proportion as before the collapse. Equation~(\ref{eq3})
does this effectively by keeping the dependence on all $\vect{q}_{i}$ other
than $\vect{q}_{k}$ the same as before the collapse.

After the collapse, $\psi$ is replaced by $\psi_{c}$ and the time development
continued as given by equation~(\ref{eq1}) until the next collapse.

When this process of time development of the wavefunction and its
occasional collapse is computer animated and displayed on screen, many
interesting effects are observed. But the overall visual effect is
what one looks for. It brings us a little closer to that much desired
``intuitive understanding''.

\section{The binding potential}

The algorithm of the previous section is general for any arbitrary hamiltonian
$H$. For a specific composite object, the interparticle binding potential
needs to be specified. Clearly, we cannot use the binding potential of a
real baseball. Even if we actually knew what it is, it would not be a
practical choice for numerical computation. So, I choose a potential that
is simple but realistic. As all bound particles near equilibrium can be
approximated to be in a harmonic (``spring'') potential, I choose a
multiparticle harmonic potential for the present computation. Each particle
of the composite is assumed to be attached by a spring to a common
center. The unextended lengths of the springs are assumed to be zero. This
results in the following hamiltonian.
\beq
H=\sum_{i=1}^{N}\frac{\vect{p}_{i}^{2}}{2m_{i}} + 
\sum_{i=1}^{N}\frac{k_{i}}{2}(\vect{q}_{i}-\vect{q}_{c})^{2}, \label{eq5}
\eeq
where $m_{i}$ is the mass and $k_{i}$ the spring constant for the $i^{\mbox{th}}$
particle. $\vect{q}_{c}$ is the position of the common center.
\begin{figure}
\hfil\includegraphics{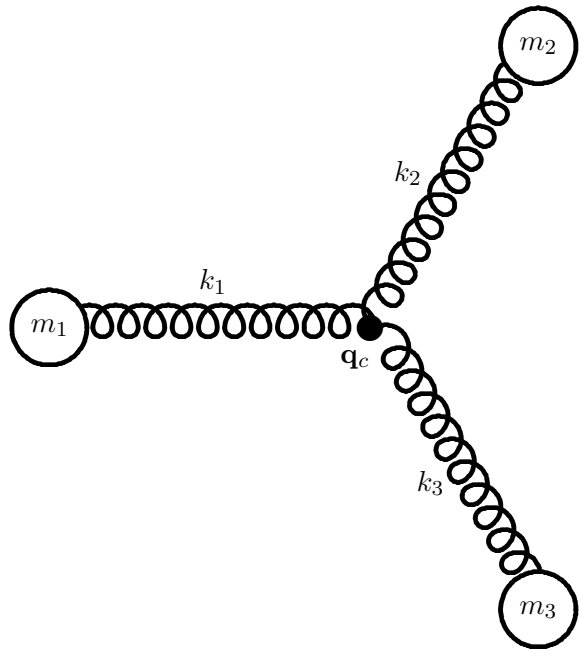}\hfil
\caption{Three particles attached by ``springs'' \label{fig1}}
\end{figure}
 The $N=3$ case of this system can be visualized as in figure~\ref{fig1}.
As the sum of the internal forces must be zero, the following
condition must be satisfied.
\beq
\sum_{i=1}^{N}k_{i}(\vect{q}_{i}-\vect{q}_{c})=0, \label{eq6}
\eeq
which gives
\beq
\vect{q}_{c}=\frac{\sum_{j=1}^{N}k_{j}\vect{q}_{j}}{\sum_{j=1}^{N}k_{j}}. \label{eq7}
\eeq
Hence,
\beq
H=\sum_{i=1}^{N}\frac{\vect{p}_{i}^{2}}{2m_{i}} + 
\sum_{i=1}^{N}\frac{k_{i}}{2}
\left(\vect{q}_{i}-\frac{\sum_{j=1}^{N}k_{j}\vect{q}_{j}}{K}\right)^{2}, \label{eq8}
\eeq
where $K=\sum_{j=1}^{N}k_{j}$.

\section{Outline of a numerical approach}
The computer screen being 2-dimensional, I shall simulate the above N-particle
composite in two dimensions. To use standard finite difference methods, the screen
space is divided into a matrix of $m$ columns and $n$ rows to produce a total of
$m\times n$ points. Although $m$ and $n$ need to be large for accuracy, practical
animation on a PC requires that they be small (less than 10 each). Hence, if this
method is used to build an animated computer game, $m$ and $n$ must be given small
values. This is not a problem as, for the purposes of a game, only qualitative aspects
need be displayed.

To solve the Schr\"{o}dinger equation, boundary conditions must be specified. There
are several possible natural choices:
\begin{enumerate}
\item Perfectly reflecting boundary conditions.
\item Perfectly absorbing boundary conditions.
\item Periodic boundary conditions.
\end{enumerate}
The perfectly reflecting
boundary produces a discontinuity at the boundary that interferes with visualization.
The perfectly absorbing boundary allows particles to go off screen, thus making
them useless for visualization. The periodic condition seems to be the best for 
visualization. It identifies the left edge to the right and the bottom edge to the
top (toroidal topology). Hence, particle current that disappears on one edge reappears
on the opposite edge.

The discrete forms for the $x$ and $y$ components of each coordinate $\vect{q}_{i}$
may be written as
\beq
q_{ix}=a_{ix}\Delta x,\;\;\;\;  q_{iy}=a_{iy}\Delta y, \label{eq9}
\eeq
where $i=1,2,\ldots ,N$, $a_{ix}=1,2,\ldots ,m$, and $a_{iy}=1,2,\ldots ,n$.
$\Delta x$ is the mesh width in the $x$ direction and $\Delta y$ is the mesh width
in the $y$ direction.

The wavefunction $\psi$, at one instant of time, is a function of all coordinates
$\vect{q}_{i}$. So, its discretized form must depend on all $a_{ix}$ and $a_{iy}$.
Thus, for numerical computation, $\psi$ is represented by an array of $2N$ dimensions
(one for each $a_{ix}$ and $a_{iy}$). In the notation of the C language it
would be: $\psi[a_{1x}][a_{1y}][a_{2x}][a_{2y}]\ldots[a_{Nx}][a_{Ny}]$. For the
special case of three particles this would be: $\psi[a_{1x}][a_{1y}][a_{2x}][a_{2y}]
[a_{3x}][a_{3y}]$. For compactness of notation I can write this as:
$\psi[a][b][c][d][e][f]$ or $\psi_{a,b,c,d,e,f}$. Then the finite 
difference form of the operation
by the hamiltonian $H$ is found from equation~(\ref{eq5}) using equation~(\ref{eq9})
and the following finite difference forms of the $\vect{p}_{i}^{2}$ operators.
\beqar
\vect{p}_{1}^{2}\psi_{a,b,c,d,e,f} & = & -\hbar^{2}\left(
\frac{\psi_{a+1,b,c,d,e,f}-2\psi_{a,b,c,d,e,f}+\psi_{a-1,b,c,d,e,f}}
{(\Delta x)^{2}}+\right. \nonumber \\
& & +\left. \frac{\psi_{a,b+1,c,d,e,f}-2\psi_{a,b,c,d,e,f}+\psi_{a,b-1,c,d,e,f}}
{(\Delta y)^{2}}\right), \nonumber \\
\vect{p}_{2}^{2}\psi_{a,b,c,d,e,f} & = & -\hbar^{2}\left(
\frac{\psi_{a,b,c+1,d,e,f}-2\psi_{a,b,c,d,e,f}+\psi_{a,b,c-1,d,e,f}}
{(\Delta x)^{2}}+\right. \nonumber \\
& & +\left. \frac{\psi_{a,b,c,d+1,e,f}-2\psi_{a,b,c,d,e,f}+\psi_{a,b,c,d-1,e,f}}
{(\Delta y)^{2}}\right), \nonumber \\
\vect{p}_{3}^{2}\psi_{a,b,c,d,e,f} & = & -\hbar^{2}\left(
\frac{\psi_{a,b,c,d,e+1,f}-2\psi_{a,b,c,d,e,f}+\psi_{a,b,c,d,e-1,f}}
{(\Delta x)^{2}}+\right. \nonumber \\
& & +\left. \frac{\psi_{a,b,c,d,e,f+1}-2\psi_{a,b,c,d,e,f}+\psi_{a,b,c,d,e,f-1}}
{(\Delta y)^{2}}\right). \label{eq10}
\eeqar
Here the most common finite difference form for second derivatives is used.
Generalizing this formula for arbitrary $N$ is tedious but straightforward.
Using equations~(\ref{eq5}), (\ref{eq9}), and (\ref{eq10}) in equation~(\ref{eq1a}), 
the wavefunction for
successive time steps can be computed. The numerical method chosen here
does not maintain normalization of $\psi$. Hence, after
each time step computation, $\psi$ must be normalized\cite{Teukolsky}.

Also after each time step computation, the screen image must be updated to
provide an animated visual effect. A visually intuitive way of doing this
for $N=3$ is to represent each particle by a primary color (red, green and
blue). Then, each position on screen (a box of size $\Delta x\times\Delta y$) 
is colored by a mix of primary colors in proportion to the probabilities of
finding the corresponding particles at that position. These probabilities
are given by the finite difference form of equation~(\ref{eq2}):
\beqar
P_{1} & = & \sum_{c,d,e,f}\psi^{*}_{a,b,c,d,e,f}\psi_{a,b,c,d,e,f}, \nonumber \\
P_{2} & = & \sum_{a,b,e,f}\psi^{*}_{a,b,c,d,e,f}\psi_{a,b,c,d,e,f}, \nonumber \\
P_{3} & = & \sum_{a,b,c,d}\psi^{*}_{a,b,c,d,e,f}\psi_{a,b,c,d,e,f}. \label{eq11}
\eeqar
To produce the effect of wavefunction collapse, one uses the mouse button click
message to trigger a collapse at the point of clicking. However, clicking the
mouse button will collapse the wavefunction for the position of just one of the
particles and that too only with a probability given by equation~(\ref{eq11}). This 
probabilistic effect can be produced using a random number generator. The wavefunction
after the collapse is given by equation~(\ref{eq3}). The function 
$\Delta_{R}(\vect{q}_{k}-\vect{q}_{0})$ in its discrete form can be chosen as
the discrete form of the Dirac delta function:
\beq
\Delta_{R}(\vect{q}_{k}-\vect{q}_{0})=\left\{\begin{array}{ll}
                                        1, & \mbox{if $a_{kx}=a_{0x}$ and $a_{ky}=a_{0y}$}\\
                                        0, & \mbox{otherwise},
                                             \end{array}
                                              \right. \label{eq12}
\eeq
where the integer values $a_{kx}$, $a_{0x}$, $a_{ky}$ and $a_{0y}$ are 
defined as in equation~(\ref{eq9}).
\section{Some results}
It has been demonstrated in an earlier publication\cite{Qduck} that
displaying quantum effects in the form of a computer game can provide a useful
visual tool for the understanding of quantum mechanics. The
formulation of the three particle case of the present problem has 
inspired another such game\cite{Qfocus}.
This game illustrates some obvious and some not-so-obvious aspects of composite
object quantum mechanics.

As expected, the wavefunction collapse leaves the undetected particles unaffected.
Also as expected, the probability profile of each particle spreads with
time\footnote{The resulting mix of the primary colors produces some rather 
unusual color effects that may interest the artists amongst us.}. What is not-so-obvious
is as follows. If we start with one particle in a collapsed state (with
no velocity), with time its probability
peak moves away from those of the other particles! As the potential function used here is
attractive, this is somewhat surprising. However, closer scrutiny can explain
this phenomenon.

Consider the standard one-particle harmonic oscillator. Higher energy eigenstates
have probability peaks farther away from the origin. This means that particles that
start off with higher momenta are likely to have their probability peaks farther away.
For the present case, a particle collapsed to its position eigenstate 
has high probabilities for large momenta and hence, large energy. This makes its
probability peak move away from the other particles.

This brings us back to the baseball problem. When the position of a single
component particle (electron) is detected, it is likely to escape due to large
momentum uncertainty. But the momentum uncertainties of the remaining particles
are virtually unaffected by this detection process.

\section{Conclusion}
For the purpose of quantum mechanical analysis of observation, macroscopic objects 
like baseballs cannot be treated as just scaled up versions of microscopic objects.
Macroscopic objects are made of a large number of microscopic objects tied together
by some forces. This composite nature of macroscopic objects gives them properties
qualitatively different from microscopic ones. The quantum description of composite
objects is in principle straightforward but computationally time consuming. However,
some qualitative properties of composites can be observed in a simplified and
approximate model implemented as a computer game. In particular, it shows why
position observations do not make the total momentum of the composite unduly uncertain.


\begin{thebibliography}{99}
\bibitem{Griffiths} Griffiths D J 1995 {\it Introduction to Quantum Mechanics},
(Prentice Hall) pp~3-5, pp~374-385
\bibitem{Jammer}  Jammer M 1974 {\it The Philosophy of Quantum Mechanics}, (J. Wiley \& Sons, New York)
\bibitem{Wigner} Wigner E P 1962 {\it The Scientist Speculates} (ed.~Good I J, Heinemann, London)   
\bibitem{Schrod} Schr\"{o}dinger E 1935 {\it Proc. Camb. Phil. Soc.} {\bf 31} 555
\bibitem{Heisenberg}  Heisenberg W 1960 {\it Physics and Beyond}
(Harper \& Rowe, New York) pp~60-62
\bibitem{Goldstein} Goldstein H, Poole C P Jr. and Safko J L 2002 {\it Classical
Mechanics} (third edition, Addison Wesley)
\bibitem{Shankar} Shankar R 1994 {\it Principles of Quantum Mechanics} (second edition, Plenum 
Press)
\bibitem{Biswas} Biswas T {\it Quantum Mechanics -- Concepts and 
Applications} (available at the URL: www.engr.newpaltz.edu/$\sim$biswast)
\bibitem{Whittaker} Whittaker E T 1937 {\it A Treatise on the Analytical Dynamics
of Particles and Rigid Bodies} (Dover, New York).
\bibitem{Biswas2} Biswas T 1994 {\it Nuovo Cimento A} {\bf 107} 863
\bibitem{Biswas3} Biswas T and Rohrlich F 1985 {\it Nuovo Cimento A} {\bf 88} 125
\bibitem{Teukolsky} Press W H, Teukolsky S A, Vetterling W T and Flannery B P 1992
{\it Numerical Recipes in C} (second edition, Cambridge University Press)
pp~851-853
\bibitem{Qduck} Biswas T 2001 {\it Computing in Science and Engineering} {\bf 3} 84.
The related game is called {\it ``Quantum Duck Hunt''} (Microsoft Windows software).
It is available at the URL: www.engr.newpaltz.edu/$\sim$biswast.
\bibitem{Qfocus} The game discussed here is called {\it ``Quantum Focus''} 
(Microsoft Windows software). It is available at the 
URL: www.engr.newpaltz.edu/$\sim$biswast.
\end{thebibliography}
\end{document}